\documentclass[useAMS,usenatbib]{mn2e}
\usepackage{graphicx}
\usepackage{listings}

\usepackage{hyperref}

\title[The orbits of outer planetary satellites using the Gaia data]
{
The orbits of outer planetary satellites using the Gaia data
}
\author[N. V. Emelyanov et al.]
{N. V. Emelyanov$^{1,2}$\thanks{E-mail:emelia@sai.msu.ru},
 M. Yu. Kovalev$^{1}$,
M. I.  Varfolomeev$^{1}$,
 \\
$^{1}$  M. V. Lomonosov Moscow State Univrsity/Sternberg astronomical institute,
        Universitetskij prospect 13,
        Moscow, 119234 Russia\\
$^{2}$Institut de m\'ecanique c\'eleste
        et de calcul des \'eph\'em\'erides -- Observatoire de Paris,
        UMR 8028 du CNRS, \\
        77 avenue Denfert-Rochereau, 75014 Paris, France \\
}

\begin{document}

\date{Accepted 2022 Xxxxxxxx XX. Received 2022 Xxxxxxxx XX; in original form 2022 Xxxxxxxx XX}

\maketitle

\label{firstpage}

\begin{abstract}
Launch of the Gaia space observatory started a new era in astrometry when the accuracy of star 
coordinates increased by thousands of times. Significant improvement of accuracy was also expected 
for the coordinates of the Solar system bodies. Gaia DR3 provided us with the data which could be 
used to test our expectations. In this work, we refine the orbits of a number of outer planetary 
satellites using both ground-based and Gaia observations. From thirteen outer satellites observed 
by Gaia, we chose six to obtain their orbits. Some specific moments in using observations of outer 
satellites made by Gaia are demonstrated. These pecularities stem from scanning motion of Gaia, 
in particular from the fact that the accuracy of observations is significantly different along and 
across the scanning direction. As expected, Gaia observations proved to be more precise than those 
made from Earth, which results in more accurate satellite ephemerides. We estimate accuracy of the 
ephemerides of considered satellites for the interval between 1996 and 2030. 
As astrometric positions published in Gaia DR3 were not corrected for the relativistic light 
deflection by the Sun, we took into account this effect, which slightly diminished the rms 
residuals. In addition, relativistic light deflection by the giant planets was estimated, which, as 
it turned out, can be neglected with the given accuracy of Gaia observations.

\end{abstract}

\begin{keywords}
Natural Satellites -- Ephemerides -- Planets and satellites: general
\end{keywords}

\section{Introduction}

There is a number of reasons why outer satellites of Jupiter, Saturn, Uranus and Neptune are 
specific objects of the Solar System. Among them is their origin which is not fully explained, 
there are only hypotheses. Eccentricities of their orbits are significant, sometimes reaching 
values as high as 0.75. Orbital planes are not connected to the equators of their planets, being 
oriented in space in a wide variety of directions. Since solar perturbations are very significant 
for these objects, their motion can be modelled only by using numerical integration.

Since 2005, we have been generating ephemerides of the outer satellites 
\citep{Emelyanov2005,EmelyanovKanter2005} which can be accessed via the MULTI-SAT ephemeris server 
\citep{Emelianov2008}. Our ephemerides are regularly updated as new observations appear, the last 
significant upgrade having been reported in \citep{Emelyanov2022}.

Accuracy of the ephemerides of the outer satellites is not high. To estimate it, methods described 
in \citep{Emelyanov2010} can be used. It turns out that, for some satellites, the accuracy is so 
bad that they can be considered to be lost. Although some of them were rediscovered 
\citep{BrozovicJacobson2017}, still there are some moons with low ephemeris accuracy, which means 
that they should be searched for and discovered again.

The models of motion are based on observations. The earliest observations were positions measured 
on photographic plates. In the last decades, CCD images have been used for this purpose. This 
improved the accuracy of measured positions, however the improvement turned out to be not so 
significant.

To measure right ascension and declination of observed satellites, coordinates of reference stars 
from star catalogues are used. Thus, errors of coordinates of stars in catalogues directly 
influence the errors of measured satellite positions. Until recently, the errors of star catalogues 
exceeded the errors of positions in CCD frames. It was these errors that determined the uncertainty 
in the coordinates of satellites.

The situation changed drastically after Gaia space observatory was launched. In addition to the 
fact that atmospheric blurring of images no longer reduced the accuracy of observations, the 
accuracy of positions of stars in Gaia catalogue has become many times better than that in earlier 
catalogues. Thus we can say about an advent of a new era in astrometry, the Gaia era.
The coming of this era raised high expectations for improvement of accuracy of coordinates of 
the Solar system bodies. These hopes were rewarded after publication of online data described by 
\citet{Tanga2022}.

We extracted observations of 31 planetary satellites from the third data release of the European 
Space Agency's Gaia mission, Gaia DR3, see Appendix \ref{sec:dataset}. The links to the data were 
taken from \citep{GaiaCollab2022, Tanga2022}. Extracted data were adapted for use with our database 
of observations of planetary satellites \citep{Arlot2009}.
Gaia observations of the outer moons of Jupiter, Saturn and Uranus were used to refine their 
orbits. In addition to orbital parameters, this allowed us to estimate accuracy of observations.

Description of Gaia observations and their pecularities is given in Section \ref{sec:observations}.
The way of fitting the orbits to observations is described in section~\ref{sec:methodology}.
Section~\ref{sec:himalia} discusses some peculiarities of Gaia observations demonstrated on the 
example of observations of the Jovian satellite J6 (Himalia). Determination of orbits and analysis 
of the O-C residuals for other satellites are given in Section~\ref{sec:othersats}. 

Estimation of the ephemerides accuracy is essential part of this work. Section~\ref{sec:accuracy} 
gives comparison of accuracy of ephemerides obtained with different sets of observations.

 When fitting orbital parameters, we took into account relativistic light deflection by the 
Sun. To find out if relativistic deflection of light in the gravitational fields of the giant 
planets should be taken into account, we calculated light deflection for different distances of 
satellites from their planets. Results of these calculations are given in 
Section~\ref{sec:lightdecl}.
In the last section, we draw some conclusions on our investigation.

\section{Description of observations}
\label{sec:observations}
To determine the orbits of the satellites, we used both ground-based observations and observations 
made by Gaia. Before proceeding to description of Gaia observations, we should note that we used 
only ground-based observations of the outer satellites made between 1996 and 2022. Although we 
have an experience of determining the orbits of outer satellites from all available ground-based 
observations \citep{Emelyanov2005}, using observations made at this limited time interval allows us 
to consider them to be of equal accuracy. Observations made before 1996 have certain systematic 
errors.

All ground-based observations were taken from NSDB database \citep{Arlot2009} which is regularly 
updated as new observations are published.

We had at our disposal Gaia observations of the outer Jovian satellites J6--J13, J17 and J18, as 
well as observations of the Saturnian satellites S9 (Phoebe) and S29 (Siarnaq), Uranian moon U17 
(Sycorax), and  Neptune's moon N2 (Nereid). The links to the data were taken from 
\citep{Tanga2022}. In this paper, we used only observations of the moons J6--J9, S29, and U17. 
Analysis of observations of the remaining outer moons observed by Gaia was postponed for future 
work.

Gaia observations are specific in that they have different accuracy along and across Gaia's 
scanning direction which results in peculiar distribution of residuals in the Gaia focal plane 
\citep{GaiaCollab2018,Tanga2022}.

Positions of the satellites observed by Gaia are grouped into series, each corresponding to the 
transit of the satellite in the Gaia focal plane as Gaia spins around its axis and scans the sky. 
Each observation is provided with the position angle $P$ of the scanning direction. Each series 
contains no more than nine positions (usually from six to nine) located at about 40 s time 
interval. The series (transits) are separated by intervals from about 106 min (the time necesary 
for Gaia to rotate at an angle equal to that between two its lines of sight) up to several days.

The positions are given as ICRS right ascensions (RA) and declinations (Dec). As mentioned above, 
the accuracy of observations is significantly different along and across the scanning direction, 
the former being much better (the errors are much smaller). To compose conditional equations for 
each observation, the residuals are calculated, that is the differences in right ascensions
$\Delta \alpha$,
and declinations $\Delta \delta$ of measured and theoretical positions. Then the values
$$
  X=\Delta \alpha \cos \delta, \;\;\; Y=\Delta \delta,
$$
are obtained, where $\delta$ is the calculated (theoretical) declination.
Thus, at the plot with the axes $X, Y$, we obtain dots corresponding to single observations. For 
observations made by Gaia, the dots form some tracks perpendicular to the scan 
direction. The scatter of dots across the scan direction is noticeably larger than that along the 
scan.

\begin{figure}
   \begin{center}
    \includegraphics[width=0.46\textwidth]{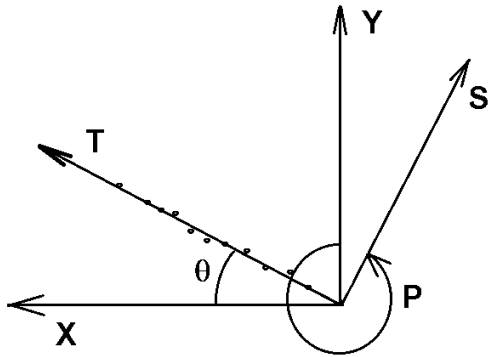}
    \end{center}
      \caption{
Spread of the residuals (dots) in the $X, Y$ coordinate system referred to the Earth equator and in 
the $T, S$ coordinate system associated with the scan direction.
              }       
\label{fig1m}
   \end{figure}

Fig~\ref{fig1m} gives an example of how residuals (dots) can be spread relative to the $X, Y$ axes. 
The origin of the reference frame corresponds to the case when observed position coincides with the 
calculated one. Let us draw axes $T, S$ so that $S$ is pointed to the scan direction, 

the axis $T$ being perpendicular to $S$.
It is seen from Fig~\ref{fig1m}, that the angle $\theta$ between the axes $X$ and $T$ is related to 
the position angle $P$ in the following way (all angles are given in degrees):
$$
   \theta = 180 - P, \mbox{ if } P \le 180, 
$$
$$
   \theta = 360 - P, \mbox{ if } P > 180.
$$

\noindent

We need to transform the residuals relative to the $X, Y$ axes into those relative to the $T, S$ 
axes (across and along the scan direction, respectively) using the given value of the position 
angle of the scan direction, $P$. Thus the residuals across the scan direction, $\Delta^{(T)}$, and 
those along the scan direction, $\Delta^{(S)}$, are obtained.
It is these residuals that are used in the conditional equations for Gaia observations.
    
When determining the satellite orbit, we should assign weights to observations corresponding to 
their errors. Each Gaia observation is provided with the covariance matrix of errors relative to 
the $X, Y$ axes. The elements of the covariance matrix should be transformed into those relative to 
the axes with maximum and minimum errors. The resulting matrix is diagonal, the diagonal elements 
being the square of the maximum error ($a$) and that of the minimum error ($b$).

Transformation of the covariance matrix was made using the method described by \citet{Deakin2005}.
The covariance matrix provided with the Gaia observations includes two components: random errors 
and systematic errors. Correspondingly, we have two versions of values for the transformed 
parameters: those obtained for random errors (we give them the subscript $r$) and the ones obtained 
for systematic errors (with the subscript $s$), i.e. $a_r$, $b_r$, $a_s$, $b_s$.
Since there are two error components, we take
$$
  a=\sqrt{a_r^2+a_s^2}, \;\;\; b=\sqrt{b_r^2+b_s^2} .
$$

Such a transformation of the standard errors provided with the Gaia data
 have been made in advance, 
before starting the process of fitting the orbits, so that the values for $P$, $a$, $b$ could be 
used as input for the program for determining the orbits. In this program, the angle $P$ is used to 
transform conditional equations relative to the corrections to the fitted parameters into the form 
where residuals along and across the scanning direction are used. The errors $a$ and $b$ are used 
to assign weights to the equations. The values for $a$ and $b$ obtained for some outer satellites 
are given in Table~\ref{tirek}.

\begin{table}
\caption{
The values for $a$ and $b$ for some outer satellites.
The value for  $a$ turns out to be constant for all observations. For $b$ we give the extreme 
values, $min(b)$ and $max(b)$. $N$ is the number of available observations.
 } 
\label{tirek} 
\begin{center}
\tabcolsep=0.6em
\begin{tabular}{lrrrr}
\hline
\hline
Satelite &  $a$, mas & $min(b)$, mas & $max(b)$, mas &  N   \\ 
\hline 
\hline
J6  Himalia  & 612.454  & 0.308 & 0.585 & 195 \\
J7  Elara    & 612.454  & 0.483 & 1.436 & 243 \\
J8  Pasiphae & 612.454  & 0.592 & 2.647 & 363 \\
J9  Sinope   & 612.453  & 1.238 & 5.320 & 169 \\
S29 Siarnaq  & 612.453  & 2.976 &22.050 & 124 \\
U17 Sycorax  & 612.453  & 3.997 &28.367 & 123 \\
\hline
\hline
\end{tabular}
\end{center}
\end{table}

\section{Methods and algorithm of determining the orbits}
\label{sec:methodology}
The trajectories of satellite motions were determined by numerical integration using initial values 
for coordinates and velocities. Initial conditions were fitted to observations by using the 
least-squares method. The detailed methodology of this process can be found in 
\citep{Emelyanov2020}. 

If there are some estimates of accuracy, weights can be assigned according to these estimates. The 
weights $w_i$ are introduced into normal equations composed as part of the least-squares method. 
For this, both left and right parts of an $i^{th}$ conditional equation are multiplied by 
$\sqrt{w_i}$. This results in that each term in the right-hand side of normal equations turns out 
to be multiplied by $w_i$. 

If error of an $i^{th}$ observation $\sigma_i$ measured in arcseconds is known, the weight 
$w_i$ can be obtained from the relation $\sqrt{w_i}=\frac{\sigma_0}{\sigma_i}$, where $\sigma_0$
is an arbitrary constant measured in arcseconds.
Detailed description of this process can be found in \citep{Emelyanov2020}. 

The dynamical model we used took into account perturbations from the Sun, other planets and 
non-sphericity of the axisymmetric body of the main planet. Positions of the Sun, planets and Earth 
were computed using the DE431 ephemerides \citep{Folkner2014}.

Attraction of the major satellites was modelled by regarding them as rings with uniform 
distribution of mass. The radii of such rings were put to be equal to the semi-major axes of the 
satellite orbits, the ring plane coinciding with that of the planet's equator. Attraction of such 
rings was taken into account by correcting for both planetary masses and the coefficients J2 and J4 
of expansion of the planet's gravitational potential. The accuracy of such approximation turned out 
to be sufficient for solving our problem. In addition, it gives better stability of the results of 
numerical integration as compared with the model where major satellites are treated as moving 
points. The formulae used in such ring-approximation can be found in \citep{Emelyanov2020}.

Since positions of the satellites obtained by Gaia were not corrected for the relativistic 
light deflection from the Sun, we took into account this effect (see more details on this issue in
Section~\ref{sec:lightdecl}).

A specific algorithm was used to determine the orbits from observations. First, equations of motion 
were integrated by using the method described by \citet{Belikov1993}, the rectangular coordinates 
of satellites being expanded as series in Chebyshev polynomials. Then, differential equations 
for partial derivatives of measured values with respect to initial conditions were integrated by 
Everhart's method \citep{Everhart1974}. These equations include the satellite coordinates 
calculated by using the Chebyshev polynomials obtained earlier. After orbital parameters were 
refined, the segments of the series representing the satellite's coordinates were saved in a file. 
Such separate integration made it possible to optimally choose integration parameters for different 
equations.

Having obtained refined parameters, we can compute (O-C) residuals, that is
the values $\Delta \alpha \cos \delta$ (residuals in right ascension)
and $\Delta \delta$ (residuals in declination).
For both right ascensions and declinations, the mean values of residuals
$mean_\alpha$ and $mean_\delta$ were computed.
Then, the root-mean-square value was computed:
$$
   \sigma=\sqrt{ \frac{1}{m} 
   \sum_{i=1}^m \left[ (\cos \delta_i \Delta \alpha_i)^2 + (\Delta \delta_i)^2 \right]  },
$$
where $m$ is the total number of observations.
This formula was used to calculate the root-mean-square residuals for ground-based observations,
$\sigma_{gb}$.

In determining the orbits, the weights $w_i=1$ were assigned to the equations for Earth-based 
observations. 

As for the Gaia observations, according to the estimates of their accuracy, the 
weights $w_i=(\sigma_{gb}/a_i)^2$ were assigned to the equations involving residuals across the scanning direction, 
while equations with residuals along the scanning direction were assigned the weights
$w_i=(\sigma_{gb}/b_i)^2$, where $a_i$ and $b_i$ are the values for $a$ and $b$ obtained for the 
$i^{th}$ observation by the method described above.

At the first step of fitting the parameters, we assumed that
$\sigma_{gb}$ is equal to 0.3 arcsec. 
For the following steps, we took the value of $\sigma_{gb}$ obtained in the previous iteration.

Statistical characteristics of residual deviations were calculated separately for both ground-based 
and Gaia observations. 
Using the residual deviations $\Delta^{(T)}_i$ and $\Delta^{(S)}_i$
for $i^{th}$ Gaia observation, corresponding root-mean-square residuals were 
calculated:
$$
   \sigma^{(T)}=\sqrt{ \frac{1}{m} 
   \sum_{i=1}^m (\Delta^{(T)}_i)^2 },
$$
$$
   \sigma^{(S)}=\sqrt{ \frac{1}{m} 
   \sum_{i=1}^m (\Delta^{(S)}_i)^2 },
$$
where $m$ is the number of Gaia observations used to determine the orbit.

Mean values of the residuals
$$
   m_{T}=\frac{1}{m} 
   \sum_{i=1}^m \Delta^{(T)}_i ,
$$
$$
   m_{S}=\frac{1}{m} 
   \sum_{i=1}^m \Delta^{(S)}_i.
$$
were also calculated.

\section{Pecularities of orbit determination from Gaia observations on the example of the satellite 
Himalia}
\label{sec:himalia}
To demonstrate specificity of Gaia observations, we use only observations of the satellite J6 
(Himalia). To determine its orbit, both ground-based and Gaia observations were used, the latter 
covering the interval between 2014 and 2017 and including 23 transits (series of observations).

\begin{figure}
   \begin{center}
    \includegraphics[width=0.46\textwidth]{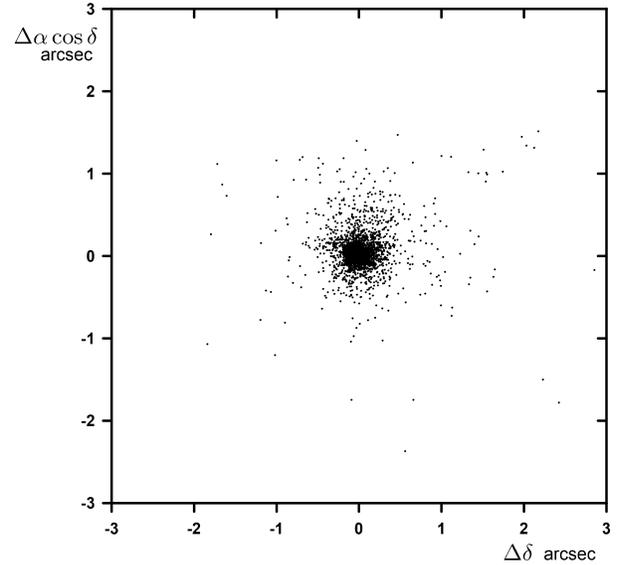}
    \end{center}
      \caption{
Residuals for ground-based observations of Himalia after fitting the orbital parameters.
              }   
\label{fig1}
   \end{figure}

\begin{figure}
   \begin{center}
    \includegraphics[width=0.46\textwidth]{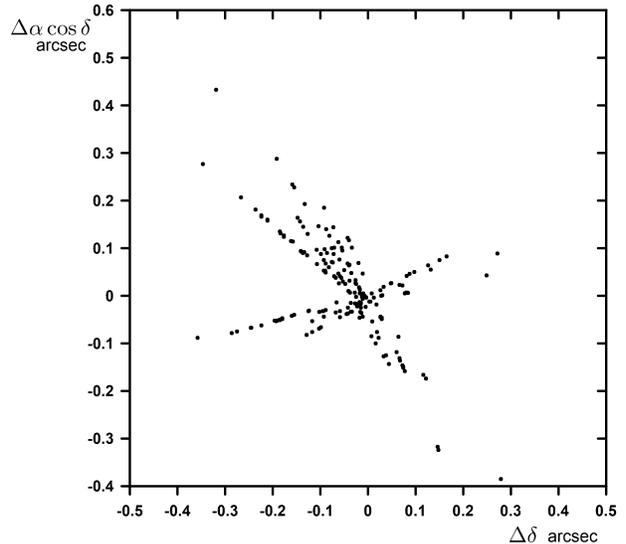}
    \end{center}
      \caption{
Residuals in the axes (X, Y) for Gaia observations of Himalia after fitting the orbital parameters.
              }   
\label{fig2}
   \end{figure}

Fig.~\ref{fig1} shows residuals for ground-based observations of Himalia. Fig.~\ref{fig2} gives 
residuals for Gaia observations of this satellite. Statistical characteristics for these residuals 
are given in Table~\ref{StatResiduJup06TW1}.

\noindent

\noindent

As  mentioned earlier, accuracy of satellite positions is significantly different along and across 
the scan direction. This leads to the need to use the measured deviations separately 
along and across the scanning direction to determine the orbit, subsequently assigning different 
weights to these measurements.

According to our methodology, it was necessary to transform deviations of 
observed positions from the modelled ones into the reference frame where axes point along and 
across the scanning direction. 
We use a reference system with an $Y$-axis pointed to the North and $X$-axis pointing to the 
direction of increasing right ascensions (to the left when viewed from the centre of the celestial 
sphere, see Fig~\ref{fig1m}).

\begin{figure}
   \begin{center}
    \includegraphics[width=0.46\textwidth]{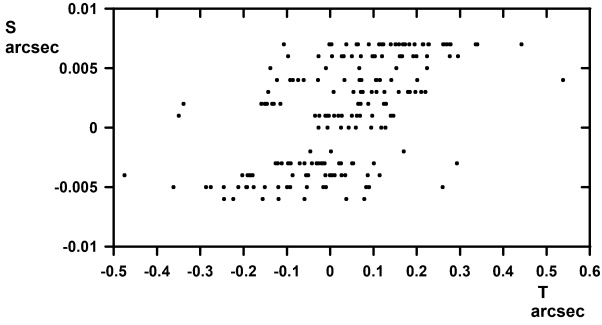}
    \end{center}
      \caption{
Residuals across and along the scanning direction for Gaia observations of Himalia.
Note that the scales for the T and S axes are different.
              }   
\label{fig4}
   \end{figure}

If there had been no errors in observations, the satellite's positions should have been located at 
the origin. The coordinate $T$ is the measure of deviation of observations across the scan 
direction, while the coordinate $S$ is that along the scan direction. Relationship between the 
coordinates $T, S$ and $X, Y$ is given by the formulae:
$$
   T= X \cos \theta + Y \sin \theta, 
$$
$$
   S= - X \sin \theta + Y \cos \theta. 
$$
Deviations computed initially in the $X, Y$ system, are transformed into those in the $T, S$ 
system. Corresponding relationships between differentials from coordinates in both systems are used 
to build conditional equations using the residuals in the $T, S$ system. The same procedure was 
used for all series of Gaia observations. Fig.~\ref{fig4} shows residuals in the $T, S$ system, 
that is residuals both across and along the scanning direction for all Gaia observations of 
Himalia. Here, the weights were assigned to the conditional equations as described above. 

\noindent

\begin{table}
\caption{
Statistical data for residuals of Himalia obtained both across and along the scan direction.
The values are given in arcsec. For ground-based observations 
two mean values are given: the first one is in right ascension, the second one in declination.
For ground-based observations the column $\sigma$ gives the value for $\sigma_{gb}$.
$N$ is the number of observations. 
 } 
\label{StatResiduJup06TW1} 
\begin{center}
\tabcolsep=0.6em
\begin{tabular}{lcrr}
\hline
\hline
Residuals & $mean$ & $\sigma$ &  N   \\ 
\hline 
\hline
Ground-based           & 0.01014, 0.04412  & 0.33320 &   3572 \\             
Gaia, across the scan  & 0.02784 & 0.15365 &    195 \\             
Gaia,  along the scan  & 0.00069 & 0.00371 &    195 \\
\hline
\hline
\end{tabular}
\end{center}
\end{table}

\section{Determination of orbits of other satellites}
\label{sec:othersats}
We also determined the orbits of the satellites J7 (Elara), J8 (Pasiphae), J9 (Sinope), S29 
(Siarnaq) and U17 (Sycorax). The orbits were fitted to both ground-based and Gaia observations.
Plots of residuals for ground-based observations of these satellites are similar to those for 
Himalia so we do not show them giving only the root-mean-square residuals.

For all satellites, calculations were performed using the same algorithm as that used for Himalia. 
Table~\ref{StatResiduOther} gives obtained estimates of the accuracy.

\begin{table}
\caption{
The mean and root-mean-square residuals for other satellites. 
For ground-based observations two mean values are given: the first one is in right ascension, 
the second one in declination.
The other columns give: the number of observations, $N$, 
the number of series (tracks) of Gaia observations, $N_{tr}$,
the minimum and maximum number of observations in the series, $(N_{min}, N_{max})$.
 } 
\label{StatResiduOther} 
\begin{center}
\tabcolsep=0.5em
\begin{tabular}{|l|c|c|r|c|}
\hline
\hline
             &         &  $\sigma$  &      &   $N_{tr}$     \\ 
Observations &  $mean$ & arcsec     &  $N$ & $(N_{min}   $  \\ 
             &  arcsec &            &      & $ -N_{max}) $  \\ 
\hline 
\hline
\multicolumn{5}{|c|}{Satellite: J7 Elara}\\
\hline
Ground-based &  0.0418, 0.07253  & 0.481  &  1975  &    -     \\
\hline
Gaia, across  &  -0.015884   & 0.145396   & 243  & 29   \\
Gaia, along   &   0.000234   & 0.003000   & 243  & (6-9) \\
\hline
\multicolumn{5}{|c|}{Satellite: J8 Pasiphae}\\
\hline
Ground-based &  0.01619, 0.11408  & 0.488   &  3521  &    -     \\
\hline
Gaia, across & -0.013267    & 0.127228   & 363  & 44   \\
Gaia, along  &  0.000288    & 0.003065   & 363  & (2-9) \\
\hline
\multicolumn{5}{|c|}{Satellite: J9 Sinope}\\
\hline
Ground-based &  0.07609, 0.12236 &  0.587   &  1384  &    -     \\
\hline
Gaia, across &  0.036790  & 0.134685   & 169  & 20   \\
Gaia, along  &  0.000506  & 0.003173   & 169  & (4-9) \\
\hline
\multicolumn{5}{|c|}{Satellite: S29 - Siarnaq}\\
\hline
Ground-based & -0.00036, 0.12340   & 0.442   &  495 &    -     \\
\hline
Gaia, across & 0.044968 & 0.128222   & 117  & 14   \\
Gaia, along  & 0.000547 & 0.009116   & 117  & (7-9) \\
\hline
\multicolumn{5}{|c|}{Satellite: U17 - Sycorax}\\
\hline
Ground-based & 0.01023, 0.07721 & 0.648   &  424 &    -     \\
\hline
Gaia, across &  -0.021410   & 0.171708   & 123  & 15   \\
Gaia, along  &   0.000645   & 0.008359   & 123  & (6-9) \\
\hline
\hline
\end{tabular}
\end{center}
\end{table}

Observations of the Uranian satellite U17 (Sycorax) are specific in the way they are distributed 
over the interval of its observations. First two published observations have been made on June 1 
and 2, 1984. Then, no observations had been made over 13-yr time interval, resuming only on 
September 6, 1997. Since that date, observations are distributed more or less uniformly over time. 
Obviously, orbital parameters of this moon are fitted to more numerous observations made since 1997. 
Using both our ephemerides and JPL ephemerides (we used Horizons System \citep{Giorgini1996} 
available at \url{https://ssd.jpl.nasa.gov/horizons/app.html}), we obtained the residuals
$\Delta \alpha \cos \delta$ and $\Delta \delta$ for two observations made in 1984 which proved to be 
suspiciously large. 
To find the reason for this, we compared the residuals for the first observation (made on June 1, 
1984) obtained with our ephemerides and those obtained with JPL ephemerides.
For the observation made on June 1, 1984, our ephemerides give
$\Delta \alpha \cos \delta$ = -4.273 arcsec,
$\Delta \delta$ = -2.657 arcsec.
With the JPL ephemerides, these values are:
$\Delta \alpha \cos \delta$ = -4.8 arcsec,
$\Delta \delta$ = -3.0 arcsec.
Small disagreement between the ephemerides is explained by differing models of motions and, 
possibly, by differing sets of observations.
At the 13-yr time interval, small differences in models can give available discrepancies between 
the ephemerides. We conclude that such big deviations of observations from theoretical positions 
are caused by the errors of observations.
For the observation made on June 2, 1984, residuals turned out to be not so big, about 1.5 arcsec.

Despite significant residuals for the 1984 observations, we left them for orbit determination, 
since they give an increase in the time interval of observations and, therefore, more reliable 
orbital parameters.

We did not reject observations with big errors, so that all published observations are included. 
Residuals proved to be random, no significant systematic shifts in residuals were found.

Obtained results demonstrate better accuracy of Gaia observations compared to that of ground-based 
observations.

We have obtained new values of the initial conditions for integrating the equations of satellite 
motions. Publication of these values here is not of interest since we use a certain dynamical model 
and a specific way to calculate the perturbing forces. Thus, the obtained initial conditions 
correspond exactly to our dynamic model. Applying them to any even slightly different model will 
certainly give other ephemeris satellite positions.

Note that the sets of observations we used to determine the orbits differ from those used 
by \citet{BrozovicJacobson2022}. Perhaps that is why the root-mean-square residuals for ground-based 
observations turned out to be larger than those obtained by JPL authors.

In total, Gaia observed 13 outer planetary satellites. We determined the orbits for only six of 
them. Although we also have observations of the Saturnian moon Phoebe, we left analysis of 
Phoebe's observations to other researchers, since the high-accuracy ephemeris of this satellite has 
been developed by \citet{Desmars2013}. Information about observations of the remaining six 
satellites is given in Table~\ref{SatAll}.

\begin{table}
\caption{
The dates and number of observations of the outer satellites observed by Gaia which were not 
analysed in this paper. ``GB obs. dates'' are the dates of the ground-based observations,
$N_{gb}$ their number,
``Gaia obs. dates'' the dates of observations by Gaia,
$N_{G}$ their number,
$N_{tr}$ the number of tracks in Gaia observations.
Note that for the satellites J10--J13 the dates and numbers of observations are given after 
observations made before 1996 were rejected.
For the two remaining satellites we give dates and number of observations.
 } 
\label{SatAll} 
\begin{center}
\tabcolsep=0.4em
\begin{tabular}{lcrcrr}
\hline
\hline
           & GB       &         &  Gaia      &         &          \\ 
Satellite  & obs.     & $N_{gb}$&  obs.      & $N_{G}$ & $N_{tr}$ \\ 
           & dates    &         &  dates     &         &          \\
\hline
\hline
J10 Lysithea & 1996.06.22 &  717       & 2014.12.14 & 136        &  16    \\
             & 2019.07.10 &            & 2017.02.24 &            &        \\
\hline
J11 Carme    & 1996.06.22 &  1767      & 2014.10.14 & 255        &  30    \\
             & 2020.08.06 &            & 2017.01.13 &            &        \\
\hline
J12 Ananke   & 1996.06.22 &  1043      & 2014.10.25 & 178        &  21    \\
             & 2020.10.17 &            & 2017.02.27 &            &        \\
\hline
J13 Leda     & 1996.06.24 &   258      & 2014.12.14 & 173        &  20    \\
             & 2018.05.19 &            & 2017.02.26 &            &        \\
\hline
J17 Callirrhoe & 1999.10.06 &   220    & 2015.12.30 &   8        &   1   \\
               & 2021.10.02 &          & 2015.12.30 &            &       \\
\hline
J18 Themisto   & 1975.09 30 &   136    & 2015.12.01 &  84        &  10   \\
               & 2021.10.02 &          & 2017.02.25 &            &       \\
\hline
\hline
\end{tabular}
\end{center}
\end{table}

\section{Estimates of the ephemeris accuracy}
\label{sec:accuracy}
The least-squares method gives us estimates of the accuracy of the determined parameters. But these 
estimates do not give an idea of the accuracy of the ephemerides. However, it is the residuals of 
the ephemerides that determine the accuracy of the model built on the basis of observations, the 
main contribution to the errors of the model being made by observational errors.
Therefore, we are more interested in the accuracy of the resulting ephemerides

To evaluate the ephemeris accuracy, we used one of the methods described by \citet{Emelyanov2010}. 
The values of the parameters were given variations using random number generator. For this, a 
covariance matrix was used, which was obtained when fitting the parameters. For each variant of 
parameters, ephemerides were computed for a number of dates. Variations of the ephemerides make it 
possible to obtain accuracy estimates. As a characteristic of accuracy, we took the 
root-mean-square value of the ephemerides variation. 1000 trials were taken for each date. More 
detailed description of the process can be found in \citep{Emelyanov2010} where comparison is also 
made of three different methods of acuracy evaluation and the reliability of the applied approach is 
shown.

\begin{figure}
   \begin{center}
    \includegraphics[width=0.46\textwidth]{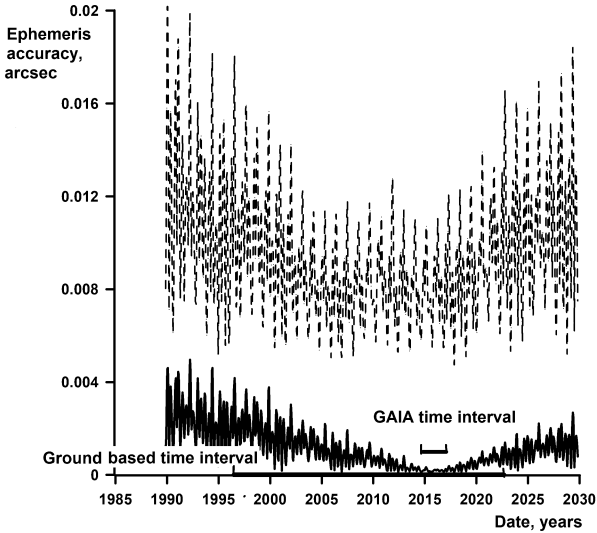}
    \end{center}
      \caption{
Estimates of the ephemeris accuracy for J6 (Himalia). The dashed line shows the estimates in 
determining the orbit without using Gaia observations, the solid line showing the estimates when 
Gaia observations are taken into consideration. Horizontal line segments show intervals of 
observations.
              }   
\label{fig06}
   \end{figure}

\begin{figure}
   \begin{center}
    \includegraphics[width=0.46\textwidth]{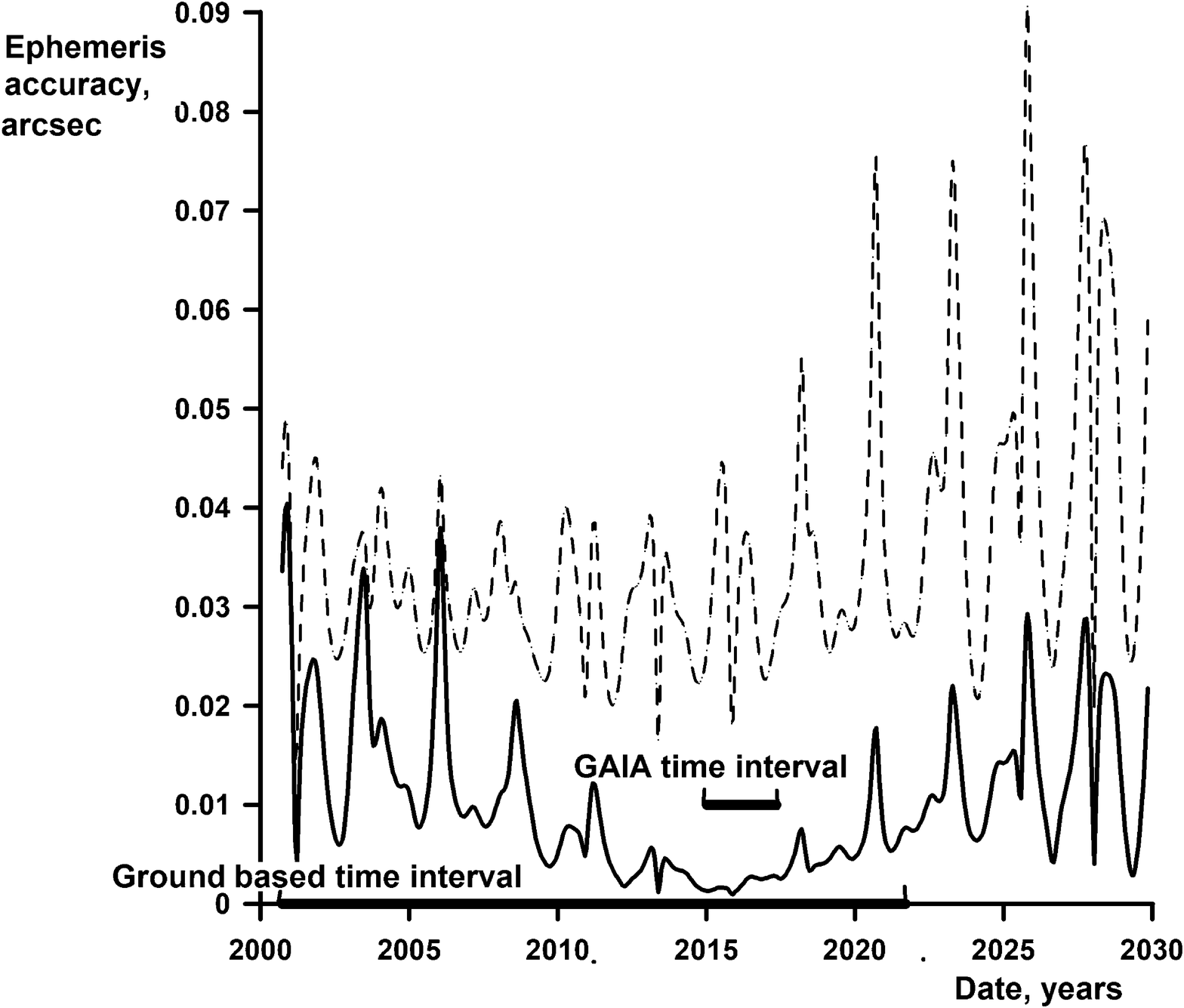}
    \end{center}
      \caption{
Estimates of the ephemeris accuracy for S29 (Siarnaq). The dashed line shows the estimates in 
determining the orbit without using Gaia observations, the solid line showing the estimates when 
Gaia observations are taken into consideration. Horizontal line segments show intervals of 
observations.
              }   
\label{fig07}
   \end{figure}

\begin{figure}
   \begin{center}
    \includegraphics[width=0.46\textwidth]{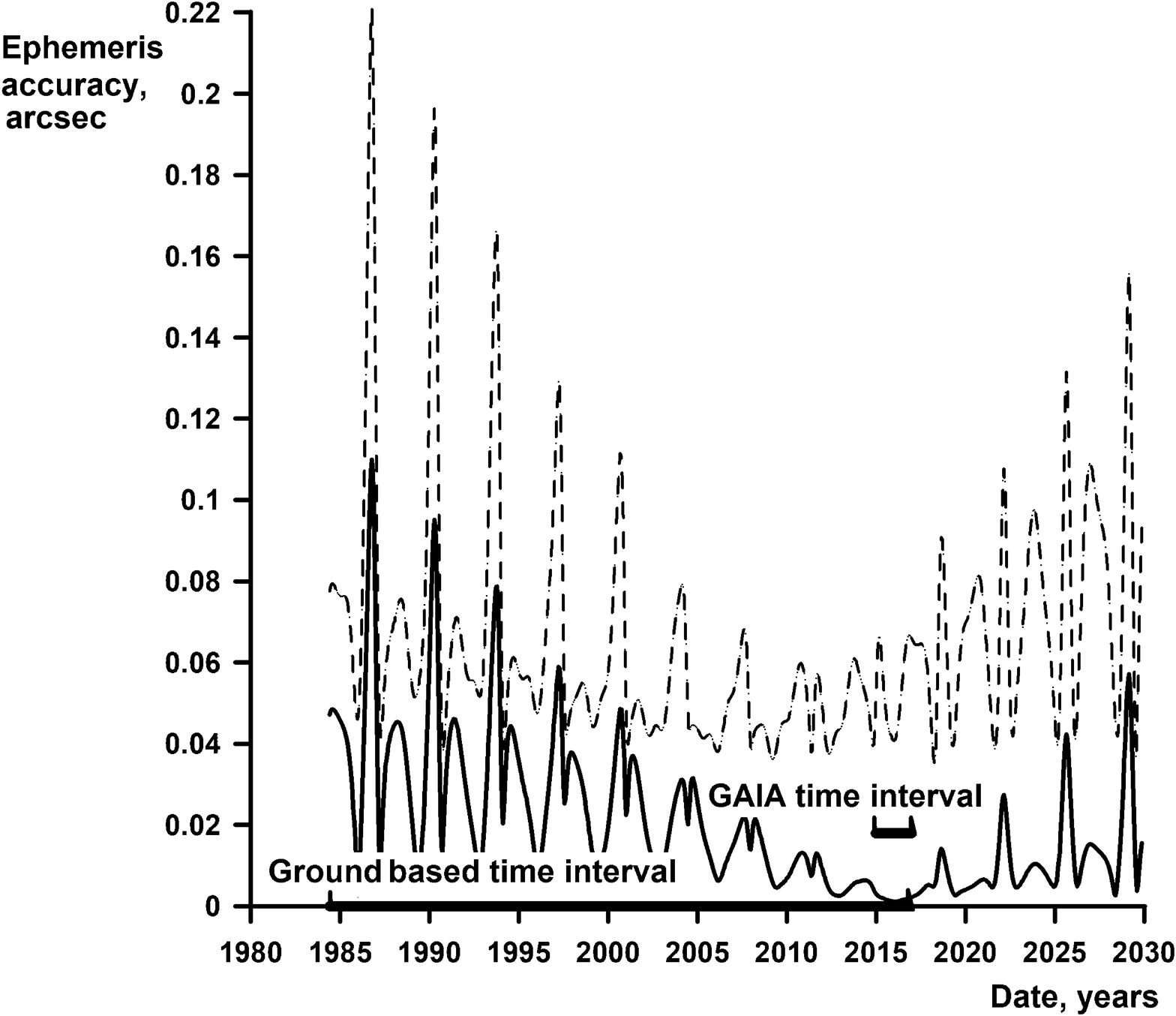}
    \end{center}
      \caption{
Estimates of the ephemeris accuracy for U17 (Sycorax). The dashed line shows the estimates in 
determining the orbit without using Gaia observations, the solid line showing the estimates when 
Gaia observations are taken into consideration. Horizontal line segments show intervals of 
observations.
              }   
\label{fig08}
   \end{figure}

The obtained estimates of the ephemeris accuracy for the satellites J6 (Himalia), S29 (Siarnaq) and 
U17 (Sycorax) are given in Fig~\ref{fig06}--\ref{fig08}, respectively. The figures also show 
intervals for both ground-based and Gaia observations.

\noindent

\noindent

\noindent

Below we give some comments on the obtained results.

Sharp fluctuations in the estimates on the plots are due to the following reasons. The ephemeris 
error is maximum along the satellite's spatial trajectory. This error increases approximately 
linearly with time from some mean date of observations. The error across the trajectory does not 
vary significantly. The velocity vector of the satellite at some moments of time is oriented along 
the plane of the sky (perpendicular to the line of sight). Thus the ephemeris error along the 
trajectory is projected on the sky plane. At other times, the satellite's velocity vector is 
directed along the observer's line of sight so that the error perpendicular to the trajectory is 
projected onto the sky plane. This error is much smaller than that along the trajectory. Therefore, 
the minima on the plots correspond to the error of the ephemerides across the trajectory, the maxima 
representing the errors along the trajectory.

Plots of the ephemeris errors prove that the errors increase almost linearly with time as the date 
moves away from the interval of observations.

It is seen from the plot for U17 (Sycorax) that the errors increase for the dates near 1985 
which is caused by significant errors of two observations made close to that date.

Comparison of error estimates for the two sets of observations, including those made by Gaia and 
without them, shows significant improvement in accuracy in the case when Gaia observations are 
taken into account.
Thus, there is an obvious progress in improving the ephemeris accuracy of the outer satellites 
when Gaia observations are used.

\section{Relativistic deflection of light}
\label{sec:lightdecl}

It is known from the general theory of relativity that light passing near a massive body is 
deflected. The angle of the deflection can be computed. Gaia DR3 documentation notes that, for 
positions of the Solar system bodies, relativistic light deflection in the gravitational field of 
the Solar System should be taken into account. This effect includes light deflection from the Sun 
and that from the giant planets.

\begin{figure}
   \begin{center}
    \includegraphics[width=0.46\textwidth]{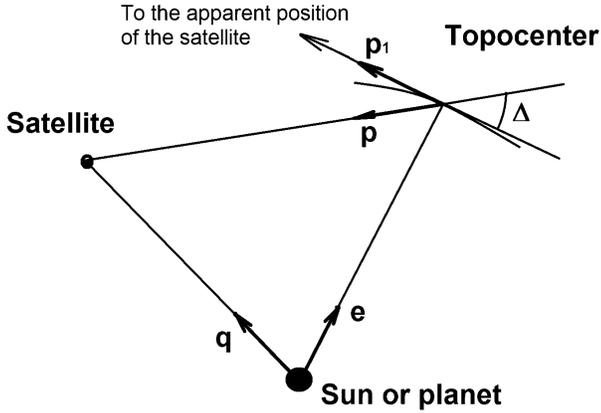}
    \end{center}
      \caption{
The arrangement of bodies to explain the relativistic deflection of light.
              }   
\label{fig09}
   \end{figure}

To calculate the deflection of light, we used the formulae given in 
\citep{ExplanatorySupplement1992} which can be used in the cases when the gravitating body (the Sun
or the planet), the observer and the observed body are located at comparable distances from one another. 
Fig~\ref{fig09} is an adaptation of the Figure 3.26.1 from the \textit{Explanatory Supplement}.
Here, a satellite is observed from the topocenter, the body deflecting the light from the satellite
being the Sun or the planet.
In both cases, the unit vectors $ {\bf e}, {\bf q}, {\bf p} $ can be obtained from ephemerides.
We use the formulae (3.26-3) and (3.26-4) from \citep{ExplanatorySupplement1992}:
$$
 {\bf p_1}= {\bf p}+\frac{g_1}{g_2}[({\bf p} {\bf q}){\bf e} - ({\bf e} {\bf p}){\bf q} ],
$$
$$
  g_1=\frac{2\mu}{c^2 E}, \;\;\; g_2= 1+ ({\bf q}{\bf e}).
$$
   For the case when deflecting body is the Sun, $\mu$ is the heliocentric
gravitational constant, $E$ the heliocentric distance of the topocenter.
   Correspondigly, if deflecting body is the planet, $\mu$ is the 
gravitational constant of the planet, $E$ is its distance from the topocenter.

For both cases, the angle $\Delta$ between the vectors ${\bf p}$ and ${\bf p_1}$, which is the 
value of the light deflection, was calculated. We denote it as $\Delta_s$ for light deflection from 
the Sun, and $\Delta_p$ for that from the planet.

The angles of light deflection were calculated only for the moments of Gaia observations.
For each satellite, we find the maximum values of $\Delta_s$ and $\Delta_p$ which we denote as
$max\{\Delta_s\}$ and $max\{\Delta_p\}$, correspondingly.
The values of $max\{\Delta_s\}$ and $max\{\Delta_p\}$ for some satellites
are given in Tables \ref{OTO} and \ref{OTOp}.

Comparing the data in both tables, one can see that the effect of deflection of light by the planet
can be neglected. However, deflection of light by the Sun cannot be ignored. Thus, the latter 
effect was taken into account in our program for fitting the orbits.

For Gaia observations of Himalia, the maximum deflection caused by the Sun gravity was 7.6 mas.
Compared to the case where fitting of parameters was made without taking into account this effect, 
the root-mean-square residuals along the scan direction diminished from 4.12 to 3.71 mas.

Note that the dates of the maximum of light deflection do not coincide with those when
the topocentric angles between the satellite and the planet were at minimum. 
This is the result of complex geometry of mutual positions of the bodies in the planetary systems.

\noindent

\begin{table}
\caption{
The events of the maximum values of light deflection by the Sun $max\{\Delta_s\}$.
} 
\label{OTO} 
\begin{center}
\tabcolsep=0.5em
\begin{tabular}{lcc}
\hline
\hline
          & $max\{\Delta_s\}$, &  Date of the event \\ 
Satellite & mas                &  (year-month-day)  \\ 
\hline 
\hline
J6 Himalia   &    7.5821        & 2016-11-26   \\
J7 Elara     &    7.6828        & 2016-11-25   \\
J8 Pasiphae  &    7.6563        & 2016-11-26   \\
J9 Sinope    &    7.7311        & 2016-11-25   \\
J10 Lysithea &    7.5822        & 2016-11-26   \\
S29 Siarnaq  &    8.8882        & 2017-01-29   \\
U17 Sycorax  &    8.8748        & 2016-02-20   \\

\hline
\end{tabular}
\end{center}
\end{table}

\begin{table}
\caption{
The events of the maximum values of light deflection by the planet $max\{\Delta_p\}$.
} 
\label{OTOp} 
\begin{center}
\tabcolsep=0.5em
\begin{tabular}{lcc}
\hline
\hline
          & $max\{\Delta_p\}$, &  Date of the event \\ 
Satellite & mas                &  (year-month-day)  \\ 
\hline 
\hline
J6 Himalia   &    0.00300     & 2015-12-01  \\
J7 Elara     &    0.00310     & 2014-10-24  \\
J8 Pasiphae  &    0.00160     & 2014-10-24  \\
J9 Sinope    &    0.00840     & 2016-06-08  \\
J10 Lysithea &    0.00140     & 2015-12-30  \\
S29 - Siarnaq&    0.00070     & 2016-04-19  \\
U17 - Sycorax&    0.00001     & 2016-12-08  \\
\hline
\end{tabular}
\end{center}
\end{table}

\section{Conclusion}
We added Gaia observations of a number of the satellites of Jupiter, Saturn and Uranus to 
ground-based observations of these satellites to refine their orbital parameters. 
The obtained results allow us to make the following conclusions.

Using Gaia observations significantly increases the ephemeris accuracy, the errors being reduced by 
several times. However, Gaia observations have a number of specific features. The main feature is 
that the accuracy of observations is different along and across direction of the scan.
This results in that the residuals form tracks along the line normal to the scan direction.
Practically, this means that, when determining the orbit, conditional equations relative fitted 
parameters should be composed with differents weights for measurements along and across the 
scan direction.

We determined the orbits of a number of outer satellites taking into account these features of Gaia 
observations. Then we analyzed the residuals obtained after fitting the parameters.
We also obtained new values of the initial conditions for integrating the equations of motion of 
these satellites. New values for the parameters coupled with our dynamical model are used to 
compute the ephemerides in the MULTI-SAT ephemeris server. New ephemerides of the satellites take 
into account results of this work.
New versions of ephemerides for other outer moons observed by Gaia and using these 
ephemerides in the MULTI-SAT server is the matter of near future.

Higher accuracy of ephemerides of outer satellites where Gaia observations were involved 
provides some new opportunities. In particular, future stellar occulations by the satellites can be 
predicted with higher degree of accuracy. More accurate ephemerides can be used to evaluate 
the accuracy of new observations. Moreover, using Gaia observations decelerates the rate of 
degrading of the ephemeris accuracy over time.

\section*{Acknowledgements}
We are grateful to the anonymous referee for a constructive report.

This work has made use of data from the European Space Agency (ESA) mission
{\it Gaia} (\url{https://www.cosmos.esa.int/gaia}), processed by the {\it Gaia}
Data Processing and Analysis Consortium (DPAC,
\url{https://www.cosmos.esa.int/web/gaia/dpac/consortium}). Funding for the DPAC
has been provided by national institutions, in particular the institutions
participating in the {\it Gaia} Multilateral Agreement.

\section*{Data Availability}
The data obtained in this work are available in the
Natural Satellites Data Center (NSDC), MSU SAI/IMCCE
(See authors affiliations) at \url{http://www.sai.msu.ru/neb/nss/indexr.htm}
and at \url{http://nsdb.imcce.fr/multisat/}.


\appendix 
\section{ADQL request for planetary satellites data extraction from Gaia DR3.}
\label{sec:dataset}
Gaia DR3 contains 158152 solar-system objects, including asteroids and planetary satellites. This work focuses only on planetary satellites, which can be selected with the following ADQL (Astronomical Data Query Language) query:\\
\begin{lstlisting}
SELECT * FROM gaiadr3.sso_observation 
WHERE source_id < -4284967286
\end{lstlisting}
providing a dataset of 5754 observations of 31 planetary satellites.

\end{document}